\documentclass[%
 reprint,
 amsmath,amssymb,
 prl,
]{revtex4-1}

\usepackage{graphicx}% Include figure files
\usepackage{dcolumn}% Align table columns on decimal point
\usepackage{bm}% bold math
\usepackage{physics}

\usepackage{hyperref}
\usepackage{color}
\definecolor{myblue}{RGB}{0,0,255}
\hypersetup{
    colorlinks,
    citecolor=myblue,
    linkcolor=myblue,
    urlcolor=myblue
}

\usepackage{color}
\usepackage{soul}
\definecolor{mycyan}{RGB}{73,8,138}
\definecolor{mypink}{cmyk}{0,0.7808,0.4429,0.1412}
\definecolor{mygreen}{RGB}{0,102,51}

\begin{document}

\title{Indefinite Causal Order in Quantum Batteries}
\author{Yuanbo Chen}
\email{chen@biom.t.u-tokyo.ac.jp}
\author{Yoshihiko Hasegawa}
\email{hasegawa@biom.t.u-tokyo.ac.jp}
\affiliation{
Department of Information and Communication Engineering, 
Graduate School of Information Science and Technology, 
The University of Tokyo, Tokyo 113-8656, Japan
}

\date{\today}

\begin{abstract}
Operations performing on quantum batteries are extended to scenarios where we no longer force the existence of definite causal order of occurrence between distinct processes. In contrast to standard theories, the so called indefinite causal order is found to have the capability of accomplishing tasks that are not possible without it. Specifically, we show how this novel class of resource comes into play in quantum batteries by first, combining two static unitary chargers into a coherently superposed one to fully charge an empty battery even if in the presence of battery's local Hamiltonian. Then we demonstrate for a non-unitary charging protocol, the indefinite causal order version charger yields a charged battery with higher energy over its classical counterpart under any conditions. We also have a finding that runs counter to our intuition which, roughly speaking, has the implication that a relatively less powerful charger guarantees a charged battery with higher energy. Finally, to reduce the cost imposed by a measurement-based protection scheme, indefinite causal order shows its potential to fulfill this goal.
\end{abstract}
\maketitle

\emph{Introduction.}--- Counter-intuitive features like quantum coherence and entanglement, which make sharp contrast from classical physics, have offered many advantages to build modern technologies. Primarily, these nonclassical properties are indispensable in quantum information processing tasks and quantum communication \cite{Nielsen:2011:QCQI}.

In the field of quantum foundations, however, the concept of causality has been extended beyond the standard quantum theory, where there exists a predefined causal order to scenarios in which no fixed causal order in advance \cite{Oreshkov:2012:NCOMMS, Chiribella:2013:PRA, Rubino:2017:SCIADV, Rubino:2021:PRR}. Other than conventional quantum theory, in this setting, the order between events exhibits nonclassical features. That is, the causal order becomes nonseparable or indefinite \cite{Procopio:2015:NCOMMS, Araujo:2014:PRL, Guerin:2016:PRL, Goswami:2018:PRL}. We call these quantum processes to have an \emph{indefinite causal order}.

Quantum battery is an emerging field of increasing interest for designing devices operating in the quantum regime capable of storing energy for later use \cite{Ferraro:2018:PRL, Andolina:2018:PRB, Ghosh:2020:PRA,Farina:2019:PRB, Rossini:2019:PRB, Andolina:2019:PRB, Le:2018:PRA}. Results developed in recent years focused on optimizing charging power \cite{JuliaFarre:2020:PRR, Andolina:2019:PRL, Campaioli:2017:PRL}, the stability of the energy cells \cite{Pirmoradian:2019:PRA, Quach:2020:PRAPPLIED}, and what role nonclassical resources plays in this context \cite{Hovhannisyan:2013:PRL}.

Most of the existing literature on quantum batteries focuses on analyzing the role of resources like entanglement and other nonclassical correlations. We ask the question, does there exist resource other than those mentioned earlier that could be beneficial to constructing quantum batteries? To offer an answer to this question, we extend operations performing on a quantum battery to scenarios where superposition of the order between different events are allowed. To be more specific, in this Letter, coherently superposed processes that exhibits nonclassical feature in the aspect of their causal order is used to accomplish tasks that can not be possible when the causal order between processes is classical.

In this Letter, we introduce a new kind of resource into quantum batteries by exploiting indefinite causal order in charging processes and the storage stage. Especially, we find that even though static driving paradigms are not able to yield a fully charged quantum battery when the battery's internal Hamiltonian is in presence, these chargers' quantum counterpart, however, unexpectedly outputs the fully charged battery. For a non-unitary charging process, we find that the upgraded charger gives us a higher-energy charged battery under any conditions. We also discover a counter-intuitive effect, which says that a relatively weaker coupling strength between charger and battery is advantageous of producing higher-quality charged batteries. Finally, indefinite causal order is also introduced to protect a fully charged battery while reducing the cost imposed by measurements. Our results shed new light on several facets of the role played by indefinite causal order in quantum batteries and open up new possibilities to further investigate non-classical resources in this context.

\emph{Charging with unitary maps.}--- The first quantum battery model under our consideration is a two-dimensional quantum system described by its local Hamiltonian.
\begin{equation}
H_B = \omega\sigma^z,
\label{eq:HB_def}
\end{equation}
where $\omega$ is a constant determined by the degree of freedom utilized for storing energy, and $\sigma^z$ is the Pauli-$z$ matrix. A charging process refers to bring the battery from a lower energy state to a higher one with reference to its local Hamiltonian. To impart energy to the battery one can achieve by rotating the (pure) state vector of the battery living in a two-dimensional Hilbert space. 

In some unitary charging processes, an external field may acts as a \emph{charger} to inject energy into the battery. A charger could also be an auxiliary system that plays a mediator's role to transfer energy to the battery, which we will consider later. Here, we characterize a \emph{charger} by expressing it as a linear combination of $\{\sigma^x,\sigma^y\}$, the other two Pauli matrices, $V=x\sigma^x+y\sigma^y$. We impose a restriction on the driving Hamiltonian such that it does not depend on time which generates a \emph{static} charging dynamics. Since we have fixed the internal Hamiltonian $H_B$, a static charging process simply means the coefficient of the driving Hamiltonian stays constant. In our consideration, the condition is whenever $x$ and $y$ do not depend on time.

During a charging process, the driving Hamiltonian $V$ together with the internal Hamiltonian of a quantum battery constitute the generator of the dynamics. The time propagator of the dynamics is then given by $U(t)=e^{-i(H_B+V)t}$, where we work in the unit of $\hbar=1$, and follow this throughout this work.

When only taking local operations into account, a collection of multiple energy cells is essentially identical to a single energy cell as we also exclude the interaction between different cells. Under this consideration, it leads us to focus on one single energy cell. Here we concentrate on charging processes where an empty cell, i.e., a qubit initialized in the state of $\ketbra{g}{g}^B$ is driven to a higher energy state, where $B$ denotes battery.

\begin{figure}[t]
    \centering
    \includegraphics[width=0.47\textwidth]{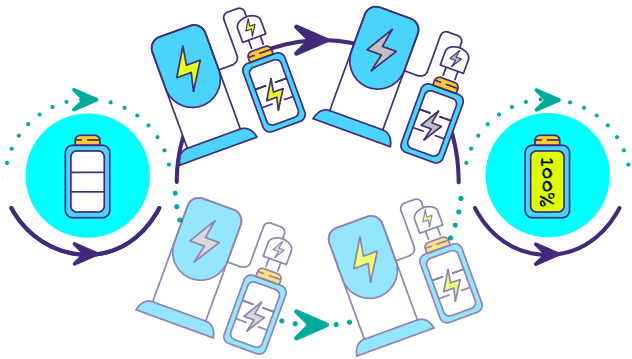}
    \caption{In a definite causal order scenario, a battery can only choose one out of two possible paths to go, however, this restriction could be removed by having a quantum switch to realize indefinite causal structure. A battery being charged by our charger experiences the superposition of causal order and benefits from this novel resource. Especially, using our superposed unitary charger, one could evolve an empty battery to a fully charged one, however, this is prohibited when the underlying causal order is classical.}
    \label{fig:ico_qb_path}
\end{figure}

As we have stated above, in this work we are going to allow processes with no fixed causal order to happen to quantum batteries, especially when several chargers are transferring energy to the battery. This idea is illustrated in Fig~\ref{fig:ico_qb_path}. The simplest non-trivial coherently controlled charger-pair needs two independent chargers to interact with the energy cell since a single charger itself represents the only possible causal order. Therefore, in what follows we have a pair of two chargers to be made of use to impart energy to an empty qubit battery.

For a pair of two chargers the temporal order of occurrence have only two different possibilites, i.e., the battery encounters $\mathcal{C}_1$ before passing $\mathcal{C}_2$ or the reverse, where we denote the $i$th charger by $\mathcal{C}_i$. A quantum switch that realizes indefinite causal order admits higher-order processes where a control qubit plays the role of determining the order of occurrence of $\mathcal{C}_1$ and $\mathcal{C}_2$. Depending on the control qubit's initial state, infinitely many configurations become possible, e.g., when the control qubit is prepared in a coherently balanced state, the battery undergoes the charging dynamics determined by $\mathcal{C}_1\mathcal{C}_2$ \emph{and} $\mathcal{C}_2\mathcal{C}_1$ in a balanced superposition. A measurement will be performed on the control qubit following the dynamical evolutions of the target system.

One can view the time evolution of charging a quantum battery as the battery passing through a quantum channel or map characterized by a Kraus decomposition, which consists of at most {$d^2$} ${N}{\times}{N}$-dimensional operators acting on the battery density operator, where $N$ and $d$ are the dimension of the target system and the environment, respectively, when the Kraus operator set is derived by tracing out the environment degrees of freedom. Any completely positive and trace-preserving (CPTP) quantum map can be represented by a set of Kraus operators that gives us information about how our target system is being transformed $\Phi(\rho) = {\sum_i}{K_i}{\rho}{K_i}^\dagger$, where ${\sum_i}{K_i}^\dagger{K_i} =\mathbb{I}$.

The quantum switch can be defined by its Kraus decomposition
%\begin{equation}
\begin{align}
W_{ij}(t)&=\ketbra{1}{1}^c\otimes K_i^0(t/2)K_j^1(t/2)\\\nonumber
&+\ketbra{0}{0}^c\otimes K_j^1(t/2)K_i^0(t/2),
\label{eq:sw_def}
\end{align}
%\end{equation}
where $K_i^0$ and $K_j^1$ are the Kraus operators for the first and the second charging process determined by two chargers respectively. Thus, the global evolution of the battery and the control qubit can be described by the following equation
\begin{equation}
\Phi_{SW}(\rho_c\otimes\rho_B) = {\sum_{ij}}W_{ij}({\rho_c\otimes\rho_B})W_{ij}^\dagger,
\label{eq:SW_evo}
\end{equation}
where $\rho_B$ and $\rho_c$ is the initial state of the battery and the control qubit of quantum switch respectively and the subscript $SW$ stands for quantum switch.

Especially in unitary evolution, the Kraus operator set consists of only a single element, the unitary operator itself. One could confirm this fact quickly by noticing that the unitary operator itself satisfies the relation $U(t)^\dagger{U(t)} =\mathbb{I}$.

Let us now investigate how indefinite causal order comes into play in the charging process dynamics. We first write down the Hamiltonian of the $i$th charger $\mathcal{C}_i$ as $V_i=x_i\sigma_x+y_i\sigma_y$. We prepare the control qubit of quantum switch in the state $\ket{+}^c=\frac{1}{\sqrt2}(\ket{0}+\ket{1})^c$. To see how the state of the battery and control qubit jointly evolve, we insert $\rho_c(0)=\ketbra{+}{+}^c$ 
and $\rho_B(0)=\ketbra{g}{g}^B$ into Eq.~\eqref{eq:SW_evo}, also noticing that the Kraus operator resulting from the quantum switch is $W(t)=\ketbra{0}{0}^c\otimes[U^0(t/2)U^1(t/2)]+\ketbra{1}{1}^c\otimes[U^1(t/2)U^0(t/2)]$, where $U^i(t)=e^{-i(H_B+V_i)t}$, we then obtain the joint density operator passing the coherently superposed channel $\rho_{cB}(t)= W(t)({\rho_c{\otimes}\rho_B})W(t)^\dagger$.

Tracing out the control qubit degrees of freedom after performing a projective measurement on the control qubit gives us the battery's reduced state, where we choose to use the basis $\{\ket{+},\ket{-}\}$, where $\ket{\pm}=\frac{1}{\sqrt2}(\ket{0}\pm\ket{1})$ for our measurement. The reduced state of the total system, hence the state of battery can be expressed as
\begin{equation}
\rho_B^\pm(t)=\textrm{Tr}_c[\ketbra{\pm}\rho_{cB}(t)\ketbra{\pm}].
\label{eq:rho_b}
\end{equation}
depending on the measurement result, the battery will evolve into different branches of future. 

We find that the superposed chargers provide us with interesting outcomes when the measurement results in the $\ket{-}$ branch. To our surprise, when a condition between the two chargers holds, after an initially empty energy cell undergoing two coherently superposed chargers we can have a fully charged battery.

To make this point clearer, imagine a battery initialized in an arbitrary pure state other than the fully charged one, it can be shown that no static charging Hamiltonian is able to drive the energy cell to the fully charged state or $\ketbra{e}{e}^B$. The impossibility of fully charging an energy cell with a static charging Hamiltonian can be observed from the fact that a pure state qubit undergoing a unitary evolution rotates about the axis of the total Hamiltonian's eigenvector. Therefore, as long as the the internal Hamiltonian is present in the charging stage, that is if it is not turned off, one fails in accomplishing the goal of fully charging a battery.

We now want to describe the condition under which an empty battery will be fully charged, also noticing that the probabilistic nature of our charging scheme originates in the process of performing a measurement on the control qubit. Unlike in a deterministic charging protocol, the \emph{probability} of getting our desirable outcome must be taken care to make the superposed-charger be useful in practice. The explicit form of the conditional state when it ends up with the $\ket{-}$ measurement result is as follows.
\begin{equation}
\rho_B^-=
\begin{pmatrix} 
{\rho_{B}^-}_{11} & {\rho_{B}^-}_{12} \\
{\rho_{B}^-}_{21} & {\rho_{B}^-}_{22}
\end{pmatrix},
\label{eq:rho_b_mat}
\end{equation}
where 
\begin{equation}
{\rho_{B}^-}_{11}=\frac{{\omega}^2(1-R)^2M^2 \textrm{sin}^2(\Omega_1t)\textrm{sin}^2(\Omega_2t)}{\Omega_1^2\Omega_2^2},
\label{eq:rho_b_11}
\end{equation}
and
\begin{align}
{\rho_{B}^-}_{22}&=\frac{(x^2_2y^2_1-2x_1x_2y_1y_2+x^2_1y^2_2)}{\Omega_1^2\Omega_2^2}\\\nonumber
&\times\textrm{sin}^2(\Omega_1t)\textrm{sin}^2(\Omega_2t),
\label{eq:rho_b_22}
\end{align}
Here, we have defined $R=x_1/x_2=y_1/y_2$ and $M^2=x^2_2+y^2_2$. The off-diagonal terms are omitted because they are not of our interest here and in fact they are forced to become zero as long as it results in a fully charged battery. Focusing on ${\rho_{B}^-}_{22}$, it turns out that when coefficients of the two chargers satisfy the following relation,
\begin{equation}
R=\frac{x_1}{x_2}=\frac{y_1}{y_2}\neq1,
\label{eq:R_condition}
\end{equation}
we have the fully charged battery with a probability that is equal to the trace of Eq.~\eqref{eq:rho_b_mat}, especially as long as Eq.~\eqref{eq:R_condition} holds, the probability is equal to ${\rho_{B}^-}_{11}$.

We turn now to investigate up to how much the probability of yielding our desirable outcome can be and the shortest time at which this happens. We find that when the relation $\frac{(1+2k)}{\Omega_2}=\frac{1}{\Omega_1}$ is satisfied, the probability becomes
\begin{equation}
p^{(k)}=1-\frac{1}{(1+2k)^2},	
\label{eq:prob_unitary}
\end{equation}
and the time it takes is
\begin{equation}
t_{min}=\frac{1}{\omega}\frac{\pi}{2},
\label{eq:t_min}
\end{equation}
where $k$ can take any positive integer.

One thing to notice about our result is that in Eq.~\eqref{eq:prob_unitary} the maximum probability that can be reached is solely determined by the relative magnitude of the charger-pair, regardless of the battery's internal Hamiltonian. Besides, this probability could be approximately taken as unity which is a good news for us. Similarly, according to Eq.~\eqref{eq:t_min} the time it takes to maximize the probability depends only on the battery itself.

\emph{Charging with non-unitary maps.}--- So far we have only considered unitary charging dynamics, despite that some non-unitary protocols have been proposed. Here we introduce indefinite causal order to assist charging energy cells whose dynamics is non-unitary. An interesting non-unitary charging protocol that uses an active equilibrium state to assist charging energy cells is proposed by the author in \cite{Barra:2019:PRL}. In this protocol, several identical auxiliary systems prepared in a Gibbs state are brought to be coupled with the energy cell sequentially and undergoes a joint evolution for some finite time.

We continue our discussion by first introducing a mediator system working as a charger, in terms of its Hamiltonian $H_C=\frac{\omega}{2}\sigma_C^z$, the battery has the same local Hamiltonian $H_B=\frac{\omega}{2}\sigma_B^z$, where we assume $\omega$ to be a positive constant. In the charging stage, the battery interacts with the charger through the interaction Hamiltonian $H_I=K(\sigma_B^+{\otimes}\sigma_C^++\sigma_B^-{\otimes}\sigma_C^-)$, where $\sigma^\pm$ is the raising (lowering) operator defined as $\sigma^x\pm i\sigma^y$. Before starting the coupling with our battery, the charger is prepared in a Gibbs state $\rho_G=e^{-\beta H_C}/\textrm{Tr}[e^{-\beta H_C}]$, where we have also set the Boltzmann constant $k_B$ to unity in addition to the Planck constant, with $\beta$ the inverse temperature.

In order to place two identical chargers into a superposed temporal order, we first derive a Kraus operator set for the dynamics of charging our battery by a single charger. Here, unlike in unitary dynamics, it may consist of more than one element at every instant of time. The dynamics of the total system is governed by the time propagator generated by local Hamiltonian terms and the interaction one, thus, $U(t)=\textrm{exp}[-i(H_B+H_c+H_I)t]$. 

Since we are exclusively focusing on the dynamics of our battery, we adopt the view of treating the charger as the environment. Noticing that the charger's initial density matrix is $\textrm{diag}(p,1-p)$, where $p$ is the population of the high-energy state of the charger, and the ratio between them is determined by $\beta$ and $K$, therefore, one Kraus operator set can be obtained as $K_{ij}(t)=\sqrt{P_j}(\bra{i}U(t)\ket{j})$, where $i, j=0$ or $1$, and $P_0=p$, $P_1=1-p$.

Having evaluated the Kraus operator set we are now ready to transform it using a quantum switch and to see what happens to the dynamics of charging our battery. Controlling by a qubit prepared in $\ket{+}^c=\frac{1}{\sqrt2}(\ket{0}+\ket{1})^c$, the output state after coming out from the superposed charger-pair, the joint density operator will be
\begin{align}
\rho_{cB}(t)&=\frac{1}{2}[(\ketbra{0}{0}^c+\ketbra{1}{1}^c)\otimes\Phi(\rho_B)(t)\\\nonumber
&+(\ketbra{0}{1}^c+\ketbra{1}{0}^c)\otimes\Delta(\rho_B)(t)],
\label{eq:map_decomp}
\end{align}
where the term $\Phi(\rho_B)$ tells us how the battery state transforms when the underlying causal order is classical, the term $\Delta(\rho_B)$ which, however, results from an indefinite causal order is of our interset. The portion of $\Delta(\rho_B)$ in the reduced state can be maximized by performing a measurement on the control qubit using the basis $\{\ket{+},\ket{-}\}$. When it results in the $\ket{-}$ outcome, $\Phi(\rho_B)$ vanishes, with only the non-classical term $\Delta(\rho_B)$ is left, therefore, the (unnormalized) reduced battery state of our interest is $\rho ^{-}_{B}(t)=\textrm{Tr}_c[\ketbra{-}\rho_{cB}(t)\ketbra{-}]=-{1/2}{\Delta(\rho_B)(t)}$.

We now begin to explore what merits $\rho ^{-}_{B}(t)$ has over its classical counterpart $\Phi(\rho_B)(t)$. Let us first evaluate the \emph{maximum} excited state population, and then focus on what the excited state population will become at the point when the \emph{measurement probability} $\textrm{Tr}[\rho ^{-}_{B}(t)]$ reaches its peak. The global highest population state is found to be $\frac{(p-1)^2}{1+2p(p-1)}$ at the time $2\pi/{\sqrt{\omega^2+K^2}}$. The second one is
\begin{equation}
f(p)=\frac{[\omega^2A(p)+K^2B(p)-C(p)\sqrt{F(p)}]G(p)}{[\omega^2D(p)+K^2E(p)+\sqrt{F(p)}]H(p)},
\label{eq:minus_max_rho00}
\end{equation}
where we have $A(p)\text{--}H(p)$ as follows, 
\begin{subequations}
\begin{align}
A(p)&=6p^4-24p^3+39p^2-23p+5\\
B(p)&=12p^4-24p^3+29p^2-15p+3\\
C(p)&=p^2+p-1\\
D(p)&=2p^2-2p+1\\
E(p)&=4p^2-4p+3\\
F(p)&=\omega^2(36p^4-72p^3+72p^2-36p+9)\\\nonumber
&+\omega^2K^2(48p^4-96p^3+100p^2-52p+14)\\\nonumber
&+K^2(48p^4-96p^3+88p^2-40p+9)\\
G(p)&=1-p\\
H(p)&=3p^2-3p+1
\end{align}
\label{eq:coeff}
\end{subequations}
and the first time at which this occurs is $\frac{4}{\sqrt{\omega^2+K^2}}\arccos\\{\frac{1}{2\sqrt{2}}\sqrt{(1/k^2H(p))(\omega^2I(p)+K^2J(p)-\sqrt{F(p)})}}$, where $I(p)$ and $J(p)$ are $6p^2-6p+3$ and $12p^2-12p+5$ respectively. When considering its classical causal order counterpart, that is in a scenario where two consecutive interactions are carried out, the best that could be achieved is described by another function, we call it $g(p)$,
\begin{equation}
g(p)=\frac{\omega^4(1-p)+2\omega^2K^2(1-p)+K^4p}{(\omega^2+K^2)^2}
\label{eq:classical_max_rho00}	
\end{equation}

We first notice that $f(0)=1$ and $f(1/2)=1/2$, based on this observation, we introduce an auxiliary function $h(p)=1-p$ to show that $f(p)\geq{g(p)}$. Since $h(p)\geq{g(p)}$ in the interval $p\in[0,1/2]$, what we should do is merely to prove that $f(p)\geq{h(p)}$. To do this, we use the fact that $h(p)$ is both concave and convex and it can also be shown that $f(p)$ is concave in the interval $p\in[0,1/2]$, hence it leads us to draw the conclusion that $f(p)\geq{g(p)}$ for $p\in[0,1/2]$, which means under \emph{any} conditions the battery will result in a higher energy state compared to what happens in classical causal order scenarios. Besides, one should realize that because the function $h(p)$ actually represents population inversion, that is what may be achieved at best by say bring many copies of charger to couple with the battery. We also stress that for our superposed charger-pair it takes merely \emph{two} runs of coupling, however, already outperforms a classical charger, for any $\omega$ and $K$. Furthermore, it exceeds the classical scenario bound that may require a large number of times of repeated coupling.

In order to gain more insight into our result, we next consider two different regimes, namely, $\omega{\gg}K$ and $K{\gg}\omega$. Let's first have a look at the classical causal order scenario. In the regime where $\omega{\gg}K$, that is when the coupling strength is comparably small, $g(p)\simeq{p}$, and $g(p)\simeq{1-p}$ for $K{\gg}\omega$, which is exactly what we have expected, since an almost vanishing weak coupling strength compared to the local energy means there is no enough exchange of energy between the charger and the battery, so a population $p$ stays where it starts from. What makes us surprised is the following \emph{counter-intuitive} result, by assuming $\omega{\gg}K$, our superposed charger behaves approximately as
\begin{equation}
f(p)\simeq{\frac{1}{1+\frac{p}{(1-p)^2}}}.
\label{eq:weak_coupling_approx}
\end{equation}
What the above result implies is that despite in a weak coupling regime where a classical counterpart charger may not even able to inject energy into the battery during two repeated interactions, our superposed charger unexpectedly behaves well and better than in strong coupling regimes, which is radically different from what we expect in classical causal order scenarios. 

\emph{Stabilizing open quantum batteries.}--- Under many circumstances, a quantum battery's state is inevitably affected when immersed in a noisy environment, especially the energy leakage induced by decoherence and relaxation is detrimental. To make a battery robust, one must tackle these unwanted effects. So far, several strategies have been proposed to stabilize charged energy cells, among which some are of \emph{passive} fashion, and \emph{active} strategies also exist. A typical example of passive strategy is by utilizing decoherence-free space properties, such as in \cite{Liu:2019:JPCC} several energy cells making up a network-like architecture prepared in a symmetry-protected dark state will become robust to a noisy environment. On the other hand, an active protecting scheme like the one described by authors in \cite{Gherardini:2020:PRR} makes use of sequential measurements to tackle the problem of storage leakage.

To implement the measurement-based protecting scheme, one needs to \emph{frequently} access the energy cell, which in practice, inevitably cause a considerable amount of cost imposed by performing measurements. It is therefore desirable to figure out ways of reducing this cost. We found that the indefinite causal order can be employed to meet this end. In what follows, we illustrate this point by considering the dissipative dynamics of a fully pre-charged energy cell and present how the indefinite causal order compensates for the loss of energy storage. Following ideas and results developed so far, here we design a coherently-controlled process building out from two identical ones, and demonstrate how the energy stored in a quantum battery is protected by this non-classical causal order process.

Suppose that the interval between two consecutive measurements becomes extremely narrow, i.e., takes the limit to zero, the Zeno dynamics will be recovered, which is then reduced to the strategy mentioned earlier. Based on the observation that even a relatively long time lapsed, it could still result in a higher energy state after undergoing a coherently-controlled process. We want to explore if there is a chance that the battery may be rescued and return to the fully charged state without relying on a Zeno dynamics. Furthermore, due to the memoryless property of Markovian dynamics, as long as the first time of rescue is fulfilled, the same could be applied again, since the battery has been reactivated to its fully charged state, the starting point of the previous cycle, thus achieving protection of our battery.

\begin{figure}[t]
    \centering
    \includegraphics[width=0.47\textwidth]{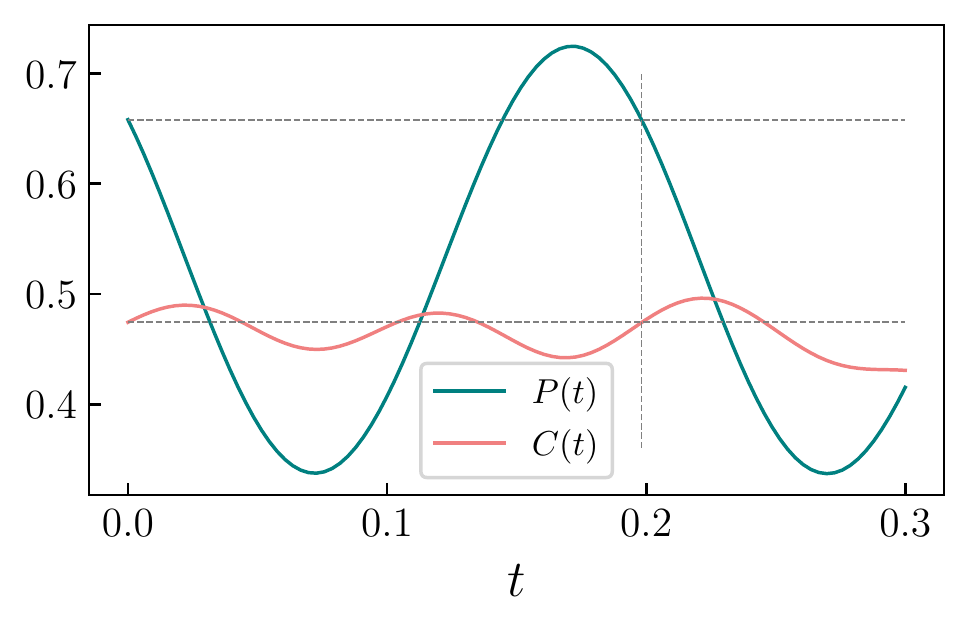}
    \caption{Evolution of the coherence and population part of the battery's density matrix, it represents what the battery would become if the control qubit was measured at that point of time. The auxiliary Hamiltonian is $H_A= 12\sigma^x+5\sigma^y$ for both two processes. One notices that $P(t)$ and $C(t)$ returns to their initial value again simultaneously after starting from a fully charged state.}
    \label{fig:ico_rescue}
\end{figure}

The model we now consider \cite{Gherardini:2020:PRR} is again a two-level system whose free evolution is captured by the Markovian master equation
\begin{equation}
\partial_t{\rho_B}=-i[H_B,\rho_B]+D[\rho_B],
\label{eq:Lindblad_eq}
\end{equation}
where $H_B$ is the internal Hamiltonian defined as $3\sigma^x+\sigma^z$ and $D[\rho_t]$ denotes the dissipator part, its explicit form is $2/3(-\{\mathcal{N},\rho_B\}+2\mathcal{N}\rho_B\mathcal{N})$, where $\{\cdot,\cdot\}$ is the anti-commutator and $\mathcal{N}=\ketbra{0}{0}$.

Before proceeding, it is important to emphasize that the variable part up to us to make different choices here is the unitary generator of the dynamics, that is we fix the dissipator and adjust only the Hermitian operators in Eq.~\eqref{eq:Lindblad_eq} by adding an auxiliary term $H_A$ to $H_B$. By employing numerical methods we explore how the reduced state of the battery changes while being driven under different conditions. For a two-state system, it suffices to focus on half of the entries in its density matrix to access all the information about the battery state. Therefore, we denote $\rho_{B11}^+(t)$ by $P(t)$ to track the population term, and use $C(t)=\abs{\rho_{B12}^+(t)}$ to represent the dynamics of coherence. We find that $H_A=12\sigma^x+5\sigma^y$ is a candidate to addressing the problem. To show this, we plot the time evolution of $P(t)$ and $C(t)$ in Fig~\ref{fig:ico_rescue}, and it can be inferred from our numerical result that the battery could be pulled back to the fully charged state if we choose to make a measurement after $t_{meas}\approx0.198$.

\emph{Conclusions.}---  We have put forward a novel non-classical resource in assisting charging qubit quantum batteries and demonstrated the usage of indefinite causal order to stabilize a fully charged quantum battery. Quite surprisingly, when the causal order of two static charger becomes non-separable, the impossibility of fully charging an empty battery in the presence of battery's local Hamiltonian is removed. By studying another non-unitary charging dynamics, we found that regardless of the properties of the battery itself and the charger, a battery will always come out from the charger in a higher-energy state when comparing to its classical counterpart. We have also shown the possibility of reducing measurement-induced cost by reactivating a quantum battery using a superposed process. The results presented here have opened up a new direction of research on quantum battery.
 
%\begin{acknowledgments}
%\end{acknowledgments}


\end{document}